\documentclass[aps,pra,superscriptaddress,twocolumn,floatfix]{revtex4-1}

\usepackage{amsmath}
\usepackage{braket}
\usepackage{graphicx}
\usepackage{epstopdf}
\usepackage{siunitx}

\newcommand{\Hamil}{\hat{\mathcal{H}}}
\newcommand{\Sp}{\hat{S}}
\newcommand{\Et}{\mathcal{E}}
\newcommand{\muB}{\mu_B}

\begin{document}

\title{Multilevel model for magnetic deflagration in nanomagnet crystals}

\author{O. Jukimenko}
\email{oleksiiiukhymenko@gmail.com}
\affiliation{Department of Physics, Ume{\aa} University, SE-901\,87 Ume{\aa},
  Sweden}

\author{M. Modestov}
\affiliation{Nordita, KTH Royal Institute of Technology and Stockholm University, SE-106\,91, Stockholm, Sweden}
\affiliation{Department of Mechanical and Aerospace Engineering, Princeton University, Princeton, NJ 08544, USA}

\author{C. M. Dion}
\email{claude.dion@umu.se}
\affiliation{Department of Physics, Ume{\aa} University, SE-901\,87 Ume{\aa},
  Sweden}

\author{M. Marklund}
\affiliation{Department of Applied Physics, Chalmers University of Technology,
  SE-412\,96 G\"{o}teborg, Sweden}

\author{V. Bychkov}
\thanks{Deceased}
\affiliation{Department of Physics, Ume{\aa} University, SE-901\,87 Ume{\aa},
  Sweden}

\begin{abstract}
We extend the existing theoretical model for determining the characteristic features of magnetic deflagration in nanomagnet crystals. 
For the first time, all energy levels are accounted for calculation of the the Zeeman energy, the deflagration velocity, and other parameters. It reduces the final temperature and significantly changes the propagation velocity of the spin-flipping front.
We also consider the effect of a strong transverse magnetic field, and show that the latter significantly modifies the spin-state structure, leading to an uncertainty concerning the activation energy of the spin flipping.  
Our front velocity prediction for a crystal of Mn$_{12}$-acetate in a longitudinal magnetic field is in much better agreement with experimental data than the previous reduced-model results.
\end{abstract}

\maketitle

\section{Introduction}

Crystals of molecular (nano-) magnets are characterized by strong magnetic anisotropy and large effective molecular spin (e.g., $S=10$ for Mn$_{12}$-acetate)~\cite{Sessoli_Nature_1993,Villain_EPL_1994}. The anisotropy implies a preferential orientation of the spin along the so-called \emph{easy} axis, leading to a considerable energy barrier between the spin-up and spin-down states of a nanomagnet. This barrier may be described as a double-well structure for the potential energy as a function of the spin projection.  When a nanomagnet crystal is placed in an external magnetic field directed along the easy axis, the states with spin along the field and against the field become stable and metastable, respectively.  Because of the barrier, the process of spontaneous quantum tunneling from the metastable to stable state is extremely slow at low temperatures~\cite{Friedman_PRL_1996,Thomas_Nature_1996,Chudnovsky_Tejada_1998,Wernsdorfer_ACP_2001,Gatteschi_ACIE_2003,Delbarco_JLTP_2005}, and the nanomagnet keeps its spin orientation upon the reversal of the magnetic field, unless spin flipping is externally induced.  These unique, superparamagnetic properties make the nanomagnets promising candidates for quantum computing and memory storage~\cite{Lis_ACSB_1980,Caneschi_JACS_1991,Papaefthymiou_PRB_1992,%
  Sessoli_JACS_1993,Bhanja_NatNano_2015}.

In nanomagnet crystals, the process of spin flipping from the metastable to stable state may happen in a form of spin avalanche known as magnetic \emph{deflagration}~\cite{Suzuki_PRL_2005,Hernandez-Minguez_PRL_2005,Garanin_PRB_2007,Villuendas_EPL_2008,Decelle_PRL_2009,Modestov_PRB_2011}. In this process, the spin flipping is triggered locally, e.g., by external heating, and the stored magnetic (Zeeman) energy is released as thermal phonon energy.  The heat is then transferred to the cold neighboring layers of the crystal by thermal conduction.  The increased temperature facilitates spin-flipping resulting in an additional release of Zeeman energy leading to a self-supporting spin-flipping front.  Such a front propagates with essentially subsonic velocity of about $\sim \SIrange[per-mode=symbol, range-units = brackets, range-phrase = \mbox{--}]{1}{15}{\meter\per\second}$.  The whole process is remarkably similar to slow combustion, also known as flame or deflagration~\cite{Law-book,Bychkov_PR_2000}, and for this reason the combustion terminology is now widely used in the studies of the magnetic spin avalanches.

Most of the work on magnetic deflagration considers the process in only one dimension, coinciding with the easy axis of the crystal. A linearized approach for weak transverse magnetic field was considered in Ref.~\cite{Friedman_PRB_98}, and more recently a few papers~\cite{Dion_PRB_2013,Subedi_PRL_2013,Velez_PRB_2014} have included the perpendicular direction and treated the whole process in a more realistic geometry. 
One problem for which the second dimension is essential concerns the magnetic instability of the deflagration front~\cite{Garanin_PRB_2013,Jukimenko_PRL_2014}.  Another important two-dimensional aspect arises from the magnetic crystal anisotropy of the crystal~\cite{Dion_PRB_2013} and the role of the transverse magnetic field on the propagation of spin avalanches~\cite{Subedi_PRL_2013,Velez_PRB_2014}.  Our previous work~\cite{Dion_PRB_2013} mainly considered the magnetic deflagration properties with respect to misalignment of the external magnetic field and the crystal easy axis.  The activation and the Zeeman energies were computed as a quantum-mechanical problem for a rather detailed system Hamiltonian.  Experimental papers~\cite{Subedi_PRL_2013,Velez_PRB_2014} provide a wide range of velocity measurements in a transverse magnetic field together with comparison to theoretical models.  However, Ref.~\cite{Velez_PRB_2014} shows a certain discrepancy between measurements and existing theory indicating the necessity for a more advanced theoretical investigation.

In the present paper, we develop two essential improvements on the existing theoretical model of magnetic deflagration. First, we demonstrate that in the presence of a strong transverse magnetic field the energy barrier structure is modified significantly, becoming three dimensional. This leads to an ambiguity in determining the activation energy, which is the key feature for calculating the front velocity and other properties of magnetic deflagration.
This problem was already recognized in \cite{Friedman_PRB_98}, using a   classical reduction of the energy barrier for weak transverse fields.
Secondly, we include all energy levels of molecular magnets in order to calculate the Zeeman energy release more accurately. This is in contrast to all previous studies, which are based on a two-level model, implying that all the spins occupy either the lowest metastable level or the ground state. Our full model yields much better agreement with the experimental measurements of the front velocity than is achieved with the previous reduced models. 

The paper is organized as follows. In the next section, we present the energy structure of the spin states of a molecular magnet in a magnetic field, including a high transverse magnetic field.  In Sec.~\ref{sec:Zeeman}, we develop a model including all energy levels and taking into account the final thermal population of spin states. Section~\ref{sec:deflagration} is devoted to the influence of the above effects on the deflagration velocity. Finally, we conclude with a discussion and brief summary of the results obtained.

\section{Energy levels in a strong transverse field}
\label{sec:energy}

Following the experimental procedure presented in Ref.~\cite{Velez_PRB_2014}, we consider a crystal of molecular magnets with the easy axis aligned in the $z$-direction.  Initially, it is fully magnetized in the opposite direction to the external magnetic field $B_z$. It is thus in a metastable state and, as will be shown in Sec.~\ref{sec:deflagration}, deflagration depends on the activation energy $E_a$ to overcome the spin reversal barrier and on the Zeeman energy $Q$ released by this spin reversal.  In addition, we consider another component of the magnetic field, $B_y$, perpendicular to the easy axis of the crystal.  We take the base temperature of the crystal as \SI{0.4}{K}, which is much lower than the gap between two consecutive levels near the ground or metastable state of the molecule~\cite{Dion_PRB_2013}.  (Note that in this paper we express energy in kelvins.)

In order to determine the activation and the Zeeman energies, we analyze the (spin) energy states of the molecular magnet with respect to an arbitrary orientation of the magnetic field, which is restricted to the $yz$-plane.  The Hamiltonian for the molecule of Mn$_{12}$-acetate can be written as~\cite{Delbarco_JLTP_2005}
\begin{equation}
  \Hamil = -D \Sp_z^2 - B \Sp_z^4  - g \muB \left( B_z \Sp_z + B_{y} \Sp_y \right) + \Hamil'.
  \label{eq:hamilton}
\end{equation}
where $D = \SI{0.548}{K}$ and $B = \SI{1.17e-3}{K}$ are the constants corresponding to the uniaxial magnetic anisotropy~\cite{Delbarco_JLTP_2005}, $g=1.93$ is the gyromagnetic factor~\cite{Sessoli_JACS_1993}, $\muB$ is the Bohr magneton, and $\Hamil'$ contains other terms such as the transverse anisotropy, intermolecular dipole interaction, and hyperfine interaction with the spin of the nuclei.  The dipolar field produced by a fully-magnetized crystal is estimated as $B_{z} \approx \SI{50}{mT}$~\cite{Garanin_PRB_2012}, while we investigate deflagration at fields $B \sim \SI{1}{T}$, hence the contribution of $\Hamil'$ to the total energy is comparatively low and will be neglected in further analysis.

\begin{figure}
\centering
\includegraphics[width=3.3in]{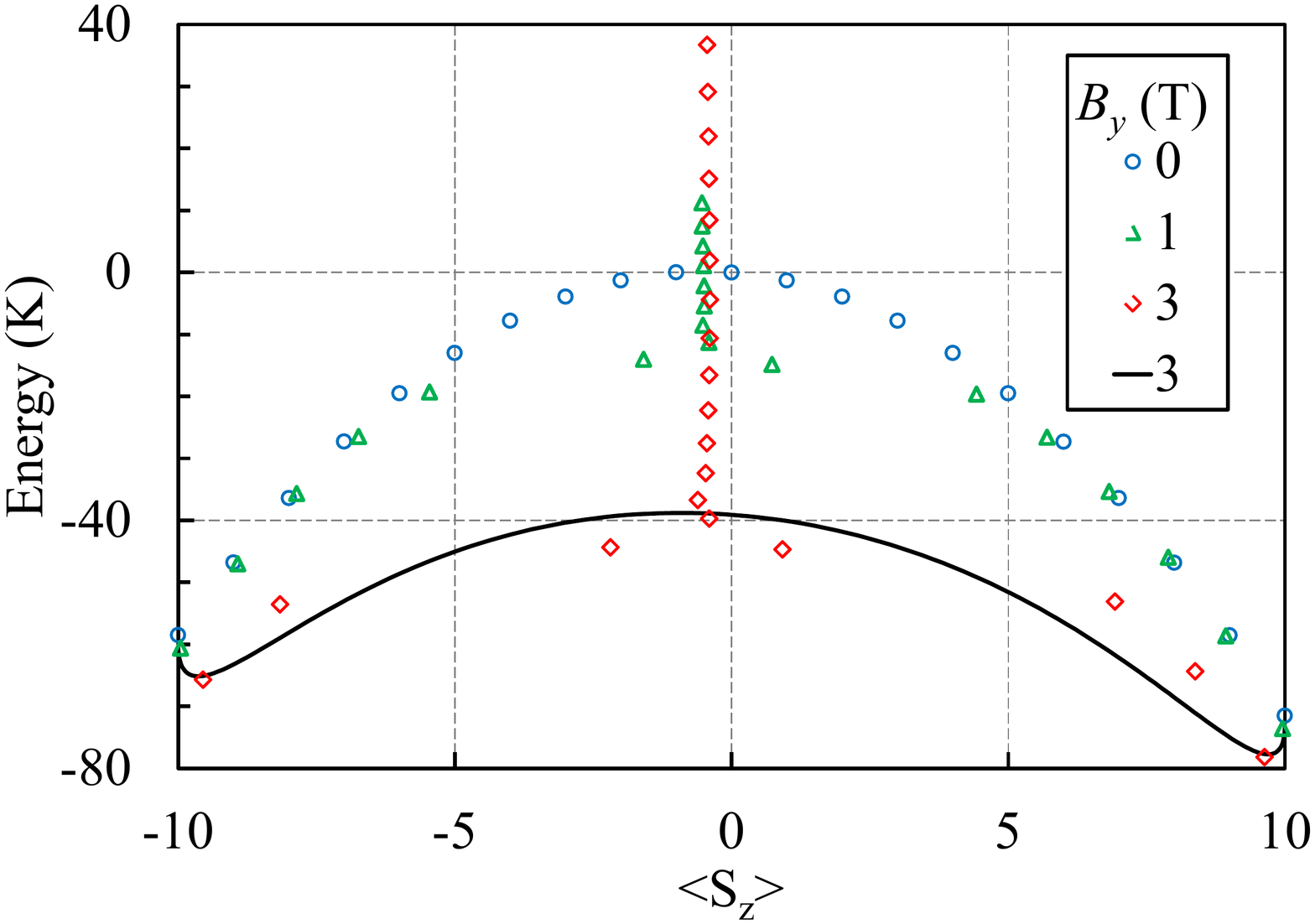}
\caption{\label{fig:states_compare}(Color online) Spin eigenstates of a single molecular magnet of Mn$_{12}$-acetate [see Hamiltonian~(\ref{eq:hamilton})] in a longitudinal magnetic field $B_{z} = \SI{0.5}{T}$, plotted according to the eigenenergy (expressed in units of K) and the projection of the spin on the $z$-axis ($\braket{\hat{S}_z}$).  Circles correspond to values in the absence of a transverse magnetic field, and triangles and diamonds to a transverse field $B_{\perp} \equiv B_y = \SI{1}{T}$ and \SI{3}{T}, respectively.  The full line corresponds to the classical approximation to the energy given by Eq.~(\ref{eq:Eclass}), for the case $B_y = \SI{3}{T}$.}
\end{figure}

The time-independent Schr\"{o}dinger equation
\begin{equation}
  \Hamil \ket{\phi_i} = E_i \ket{\phi_i},
  \label{eq:eigen}
\end{equation}
with $i = -S, \ldots, S$, can be solved numerically for by diagonalization of Hamiltonian~(\ref{eq:hamilton}) in matrix form, for different longitudinal ($B_z$) and transverse ($B_{\perp} = B_y$) fields.  As we showed in Ref.~\cite{Dion_PRB_2013}, in the presence of a small external transverse magnetic field, the actual states $\ket{\phi_i}$ are close to the eigenstates of $\Sp_z$, such that the label $i$ can be associated to the magnetic quantum number $M_z$.  The presence of a strong transverse field modifies this picture substantially, as can be seen in Fig.~\ref{fig:states_compare}, where the plot of the energy as a function of the projection of the spin on the $z$-axis shows an abundance of states with $\braket{\Sp_z} \approx 0$. Nevertheless, the ground and metastable states are not significantly affected.

However, for the calculation of the activation energy, it doesn't appear that the previous approach~\cite{Dion_PRB_2013} of associating the energy barrier to spin reversal with the highest energy eigenvalue holds in the presence of a strong transverse field.  Plotting the energy of the states $\ket{\phi_i}$ as a function of both spin projections $\braket{\Sp_y}$ and $\braket{\Sp_z}$, Fig.~\ref{fig:states_compare},
\begin{figure}
\centering
\includegraphics[width=3.3in]{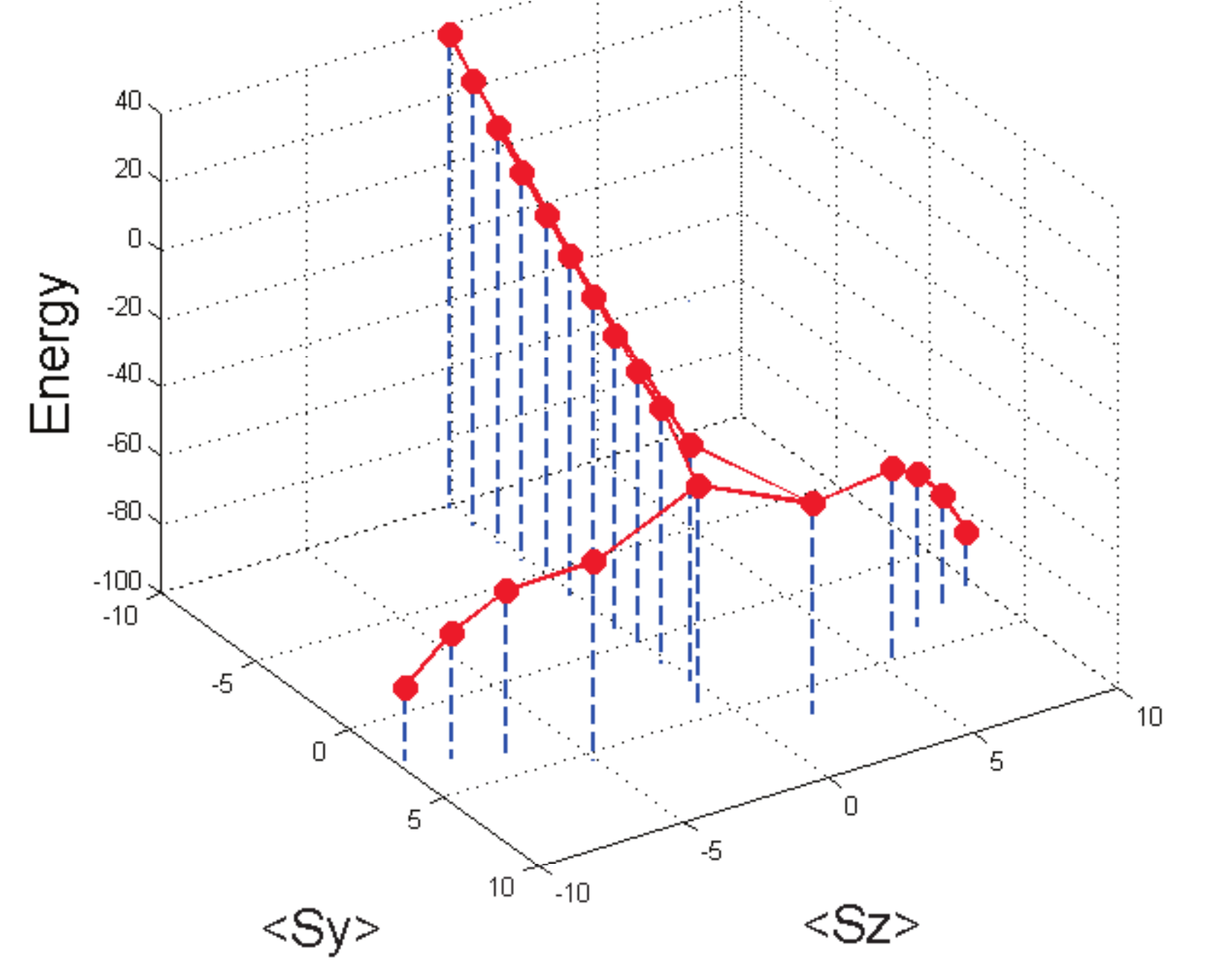}
\caption{\label{fig:states3D}(Color online) Energies of the eigenvalues calculated for Hamiltonian (\ref{eq:hamilton}) for a transverse field $B_{\perp} = B_y = \SI{3}{T}$ and a longitudinal field $B_z= \SI{0.2}{T}$. The $2S+1=21$ eigenstates are represented by (red) dots. Dashed (blue) lines locate the eigenstates on the plane.  Solid (red) lines correspond to the most probable transition between states due to the spin-phonon coupling.}
\end{figure}
one can imagine the spin reversing from $\braket{\Sp_z} \approx -S$ to $\braket{\Sp_z} \approx S$, while maintaining $\braket{\Sp_y} \approx 0$.  To test this hypothesis, considering that the spin-phonon coupling operator can be written as~\cite{Koloskova_FTT_1963}
\begin{equation}
\sum_{\alpha,\beta} c_{\alpha \beta} \Sp_\alpha \Sp_\beta,
\end{equation}
with $\alpha,\beta \in \{x,y,z\}$, we have calculated the couplings $\braket{\phi_j | \Sp_\alpha \Sp_\beta | \phi_i}$  and indicate by the solid (red) line in Fig.~\ref{fig:states3D} the strongest couplings for each state.  We clearly see that phonons can bring the system from the metastable to the ground state following the lowest energy path along $\braket{\Sp_z}$.  Unfortunately, apart for a visual inspection, there are no simple criteria that would allow to determine which of the states to use to calculate the barrier to spin reversal.
Friedman in~\cite{Friedman_PRB_98} suggests to select the highest energy state using an arbitrarily-tuned value, tunnel-splitting criterion. However, the result depends on the tuned value and is not universal.

To simplify the computations, we thus use a classical model for the spin, with energy given by the classical analogue of Hamiltonian~(\ref{eq:hamilton}),
\begin{align}
  E_\mathrm{class} &= -D S^2 \cos^2 \alpha - B S^4 \cos^4 \alpha \nonumber \\
  & \quad - g \mu_B S \left( B_z \cos \alpha + B_y \sin \alpha
  \right),
\label{eq:Eclass}
\end{align}
with $\alpha$ the angle between the spin vector $\mathbf{S}$ and the $z$-axis.  During the magnetic avalanche, we consider that the spin will move from the metastable to the ground state using the path of least resistance (lowest energy), as a function of the angle $\alpha$, as illustrated in Fig.~\ref{fig:states_compare}. We calculate numerically the extrema of $E_{\mathrm{class}}$ as a function of $\alpha$, and assign the local minimum $E_{\mathrm{meta}}$ around $\alpha \approx \pi$ to the metastable state, and the maximum $E_{\mathrm{max}}$ to the energy barrier.  We then calculate the activation energy as
\begin{equation}
E_a= E_{\mathrm{max}} - E_{\mathrm{meta}}.
\end{equation}

\section{Generalized Zeeman energy accounting for all spin states}
\label{sec:Zeeman}

\begin{figure}
\centering
\includegraphics[width=3.3in]{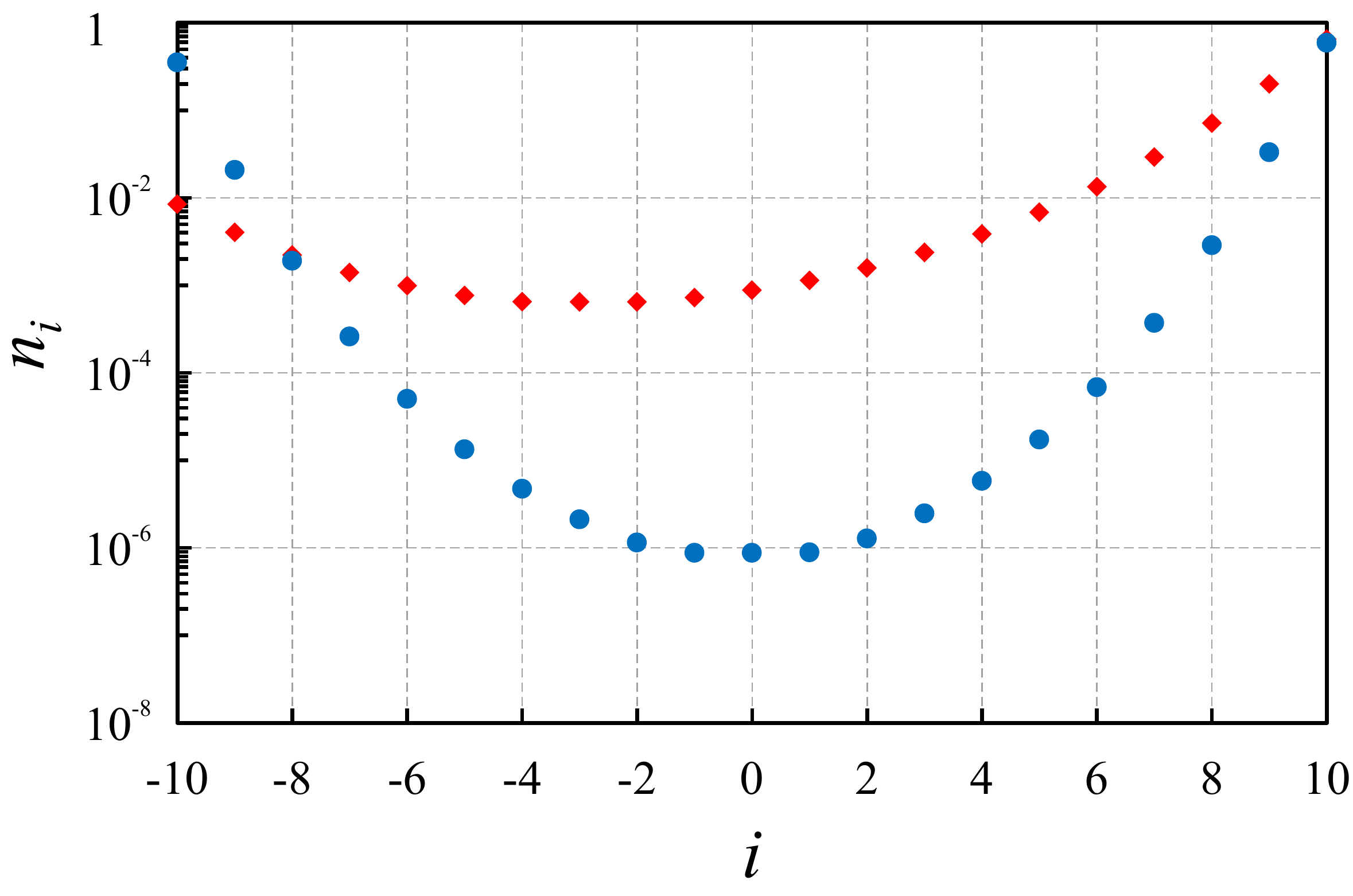}
\caption{\label{fig:states}(Color online) Relative population distribution for molecules of Mn$_{12}$-acetate. Blue circles represent the low-field $B_z = \SI{0.1}{K}$ case, with a final temperature $T_f = \SI{5.05}{K}$; most of the molecules are found in the extreme states, $n_{-10}+n_{10}=0.92$. Red diamonds represent the high-field $B_z = \SI{2.5}{T}$ case, with a final temperature $T_f= \SI{15.02}{K}$; here $n_{-10}+n_{10}=0.66$.}
\end{figure}

Having determined in Sec.~\ref{sec:energy} the activation energy in the presence of a strong transverse magnetic field, let us now focus on the Zeeman energy released by the spin flip. We designate the relative occupation (population) of each state as $n_i$, where $i=-S, \ldots, S$, representing the fraction of molecular magnets in state $i$, under the normalization condition
\begin{equation}
  \sum_{i=-S}^S n_i = 1.
  \label{norm_1}
\end{equation}
As previously, we number the states in increasing order of their spin projection along the $z$ direction, i.e., $n_{-S}$ corresponds to the fraction of molecules in the metastable state and $n_{S}$ corresponds to the fraction of molecules in the ground state.  We assume that the occupation of the energy levels is consistent with the thermal Boltzmann factor, $n_i^{\mathrm{eq}} \propto e^{-E_i/T}$. Therefore, the relative population of the $i$th level is given by
\begin{equation}
  n_i^{\mathrm{eq}}=\frac{1}{Z} e^{-E_i/T}
  \label{density_1}
\end{equation}
with 
\begin{equation}
  Z = \sum_{j=-S}^S e^{-E_j/T}
\end{equation}
the partition function and $T$ the temperature.  According to Eq.~(\ref{density_1}) the equilibrium relative populations depend on both the temperature $T$ and the external magnetic field $\mathbf{B}$, as the state energy $E_i$ depends on $\mathbf{B}$, see Hamiltonian~(\ref{eq:hamilton}).  In Fig.~\ref{fig:states} we present equilibrium state populations, determined by Eq.~(\ref{density_1}) for two values of external magnetic field.
As expected, the extreme states with $\braket{\Sp_z} \approx -S$ ($n_{-S}$) and $\braket{\Sp_z} \approx S$ ($n_{S}$) have the highest occupation numbers.  However, for a stronger magnetic field, and consequently a higher final temperature, we notice that a considerable fraction of the molecular magnets are found in the other levels.  In particular, at low magnetic field (blue circles in Fig.~\ref{fig:states}) more than 90\% of all molecules are concentrated on the two extreme levels (metastable and ground), while for $B_z= \SI{2.5}{T}$ (red diamonds in Fig.~\ref{fig:states}) this fraction decreases to 66\%.  This is due to the higher burnt temperature.  Consequently, the model used in previous studies (see, e.g., Refs.~\cite{Garanin_PRB_2007,Modestov_PRB_2011,Dion_PRB_2013}), which considered only two levels (metastable and ground), would not account for a large portion of the spin population for strong transverse fields.

In order to find equilibrium relative population and build Fig.~\ref{fig:states} from Eq.~(\ref{density_1}), it is necessary to know the final temperature of the magnetic deflagration process.  For this purpose, a more rigorous analysis of the energy balance is required.  The total energy of the system consists of thermal phonon and potential magnetic energies.  The phonon energy depends on temperature as $\Et\left(T\right)=CT^4$, where the heat capacity for Mn$_{12}$-acetate is $C\approx\SI{0.001}{K^{-3}}$
The magnetic energy of the system is defined as the sum of all energy states weighted with the corresponding relative populations,
\begin{equation}
  E_{\mathrm{mag}} \equiv \sum_{i=-S}^S E_i n_i.
\end{equation}
The total energy of the system is conserved, so before and after the deflagration front we have
\begin{equation}
  CT^4_0+n^{0}_{-S}E_{-S} = CT^4_f + \sum_{i=-S}^S E_i n_i,
  \label{totenergy}
\end{equation}
where $T_0$ is the initial crystal temperature, with all the molecules assumed to be initially in the metastable state, $n^{0}_{-S}=1$.  The latter condition can easily be fulfilled experimentally.  
Equation~(\ref{totenergy}) neglects heat exchanges with the external media, as typically the magnetic deflagration process is much faster than thermal relaxation~\cite{McHugh_PRB_2009,Velez_PRB_2014}.
The initial temperature is typically rather small, $T_0\sim \SI{0.4}{K}$, and has a negligible effect on the final result.  We solve Eq.~(\ref{totenergy}) for $T_f$ together with Eqs.~(\ref{norm_1}) and (\ref{density_1}).  Having found $T_f$, we substitute its value into Eq.~(\ref{density_1}) and obtain Fig.~\ref{fig:states}.

It is of interest to compare the above ``full'' model to the previous one using only the two extreme levels.  For the latter case, the relative populations of the metastable and ground levels reduce to
\begin{equation}
  \begin{aligned}
    n_{-S}^{*} &= \frac{1}{1+e^{Q^*/T^*}}, &  n_{S}^{*} &= 1-n_{-S}^{*}.
  \end{aligned}
  \label{density_old}
\end{equation}
Here and below, we designate by the superscript $^*$ variables within the two-level model; $Q^*$ is the Zeeman energy release, which in this case is the difference in energy of the two states, $Q^* = E_{-S}-E_{S}$. Substituting Eq.~(\ref{density_old}) into Eq.~(\ref{totenergy}), one can compute the final temperature for the two-level model, $T_f^*$. For the full model, the total energy release depends, strictly speaking, on the occupation of all the levels.  We designate the effective Zeeman energy $Q_{\mathrm{eff}}$ as difference in magnetic potential energy before and after deflagration,
\begin{equation}
  Q_{\mathrm{eff}} \equiv E_{-S} - E_{\mathrm{mag}}(T_f).
  \label{zeeman_eff}
\end{equation}

\begin{figure}
\centering
\includegraphics[width=3.3in]{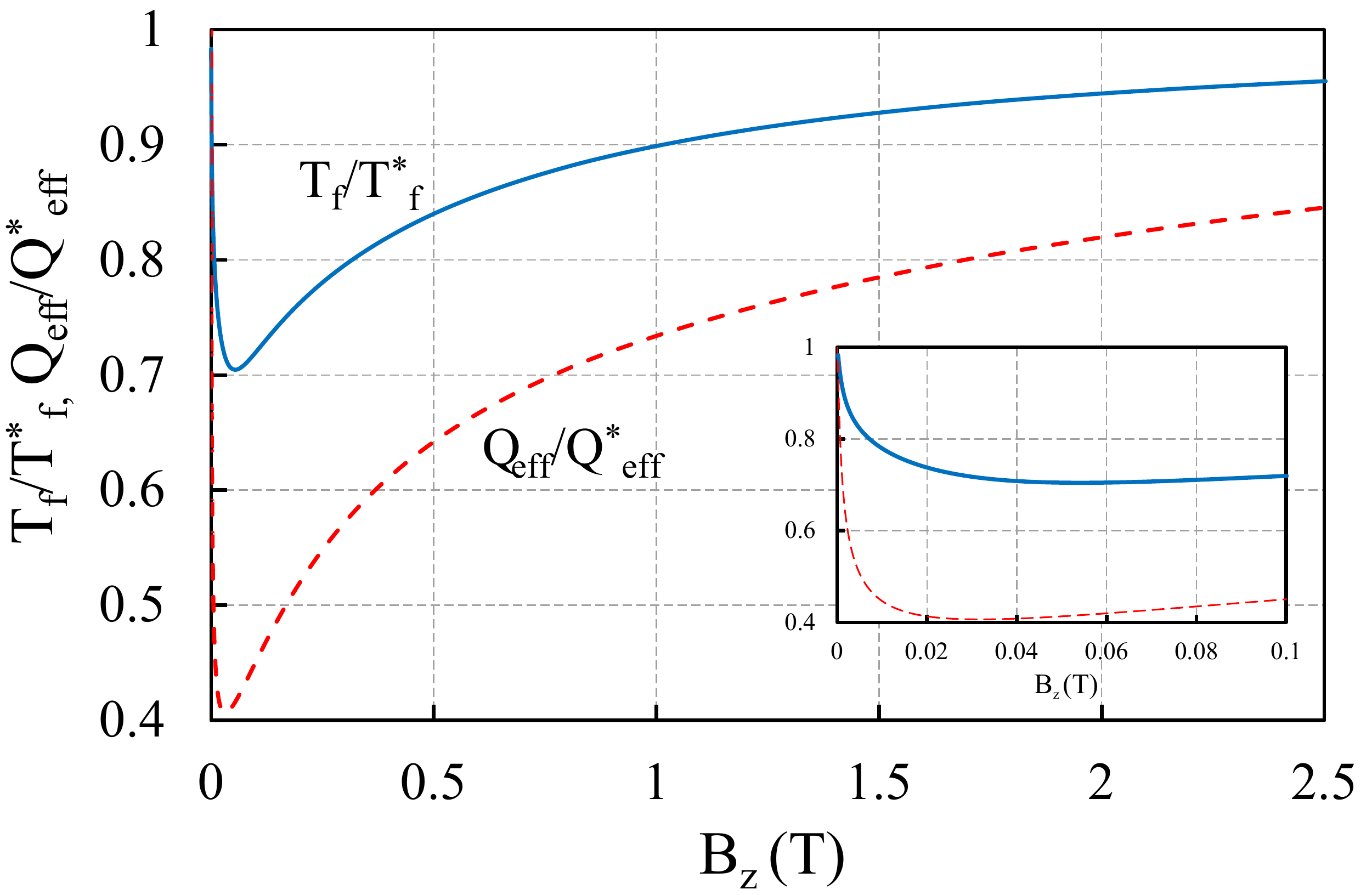}
\caption{\label{rel_par}(Color online) Ratio of the final temperature (full blue line) and effective Zeeman energy (dashed red line) calculated for the full and two-level models, as a function of the longitudinal magnetic field $B_z$. The initial temperature is $T_0=\SI{0.4}{K}$ and $B_y = 0$. The inset shows the region of weak field.}
\end{figure}

An alternative but equivalent way to compute $Q_{\mathrm{eff}}$ follows from Eq.~(\ref{totenergy}) by taking the difference of the final and initial thermal energies, $Q_{\mathrm{eff}}=\Et(T_f)-\Et(T_0)$.  Here, we obtain an important distinction between the two-level and full models.  In the first case, the Zeeman energy is purely determined from the Hamiltonian, itself linearly dependent on the magnetic field, while for the full model the Zeeman energy also depends on the temperature which, in turn, has a complicated dependence on the magnetic field.  A comparison of the final temperature $T_f$ and the effective Zeeman energy $Q_{\mathrm{eff}}$ between the full and two-level models is presented in Fig.~\ref{rel_par}.  
We see that taking under consideration all energy levels decreases the final temperature as well as the energy released in the system compared to the two-level model. A ratio close to one of both models is obtained only for very weak fields.  At low field, the assumptions of the two-level model work well since two conditions are met: (i) only the metastable and the ground states are occupied; (ii) the energy gap between the higher levels and the extreme levels is much larger than the temperature of the system.  Similarly, the spacing between the stable state and the next adjacent state increases with the longitudinal magnetic field, such that the simplified model is again a good approximation for strong fields, where only the ground state is occupied in the final configuration, even though the final temperature is higher.  The thermally-driven relaxation rate is described by the Arrhenius equation $\propto \exp(-E_a/T)$, and thus very sensitive to any change in temperature.  In the next section, we will compute the deflagration velocity for the full model and discuss comparisons with experimental data and the two-level model.

\section{\label{sec:deflagration}Magnetic deflagration front velocity}

The time evolution of the energy during the magnetic deflagration is given by
\begin{equation}
  \frac{\partial \Et}{\partial t}=\nabla\cdot(\kappa\nabla \Et) - \frac{\partial E_{\mathrm{mag}}}{\partial t},  
  \label{heat}
\end{equation}
where $\kappa$ is the thermal conductivity and the last term represents a heat source due to the Zeeman energy release.  
We follow the usual assumption that the thermal conductivity is a function of temperature, $\kappa=\kappa_0 T^{-\beta}$, where $\kappa_0$ and $\beta$ are constants, although with uncertain values.  $\kappa_0$ is usually estimated by fitting theoretical results to experiments.  
To the best of our knowledge, the exponent $\beta$ has not been measured in experiments so far, while theoretically it varies within a wide range, $\beta=-13/3 \ldots 13/3$~\cite{Velez_PRB_2014}.

It is more convenient to work in the reference frame of the deflagration front.  More specifically, we consider a front moving in the negative $z$ direction with constant velocity $U_f$.  The time dependence is eliminated by setting $f(z,t) = f(z+U_f t)$ and Eq.~(\ref{heat}) can be integrated as
\begin{equation}
  \frac{\kappa}{U_f}\frac{d\Et}{dz} = \Et-\Et_0+E_{\mathrm{mag}}-E_{-10},
  \label{heat2}
\end{equation}
where $\Et_0=\Et(T_0)$.  Now we have to specify how $E_{\mathrm{mag}}$ changes within the front.  Strictly speaking, one should consider the dynamics of all $2S+1$ states, which is quite a complicated problem.  Instead of this direct approach, we investigate the evolution of the metastable level ($n_{-10}$) only.  At the front, its relative population changes from the initial $n_{-10}=1$ in the ``unburnt'' region to the final $n_{-10}=n^{\mathrm{eq}}_{-10}$ [given by Eq.~(\ref{density_1})] in the ``burnt'' region.  Within the front, it is described by the Arrhenius law
\begin{equation}
\frac{1}{U_f}\frac{d n_{-10}}{dz} = -\Gamma_0 e^{-E_a/T} \left( n_{-10}-n_{-10}^{\mathrm{eq}} \right), 
  \label{con}
\end{equation}
where the prefactor $\Gamma_0$ is a constant and the exponential stands for the relaxation over an activation threshold $E_a$.  The activation energy $E_a$ is determined as the distance from the metastable level to the maximum of the parabola depicted in Fig.~\ref{fig:states_compare}.  We neglect here any tunneling effect~\cite{Chudnovsky_Tejada_1998,Delbarco_JLTP_2005}.
In addition, we assume that the relaxed molecules at every point of the front are distributed according to the equilibrium occupancy, Eq.~(\ref{density_1}).  In this case, $E_{\mathrm{mag}}$ is given by
\begin{equation}
E_{\mathrm{mag}} = n_{-10}E_{-10}+\frac{1 - n_{-10}}{1 - n^{\mathrm{eq}}_{-10}}\sum\limits_{i=-S+1}^S n^{\mathrm{eq}}_i E_i.
  \label{totenergy2}
\end{equation}
Finally, we rewrite the energy equation~(\ref{heat2}) in terms of the temperature as
\begin{equation}
  \frac{\kappa_0}{U_f}T^{\alpha-\beta}\frac{dT}{dz} = T^{\alpha+1}_0-T^{\alpha+1} + \frac{E_{\mathrm{mag}}-E^0_{-10}}{(\alpha+1)C}.
  \label{heat3}
\end{equation}
Equations~(\ref{con})--(\ref{heat3}) form a complete system which describes the internal structure of the magnetic deflagration front. The front velocity $U_f$ corresponds to an eigenvalue of the stationary problem. Following the numerical technique of Ref.~\cite{Modestov_PRB_2011}, we integrate the system Eqs.~(\ref{con})--(\ref{heat3}) and find the dimensionless eigenvalue $\Lambda\equiv\Gamma_0\kappa_0 T_f^{-\beta}/U_f^2$.

As stated above, the coefficients $\kappa_0$ and $\Gamma_0$ cannot be uniquely defined and are used as fitting parameters. Actually, the product $\kappa_0\Gamma_0$ is a multiplicative coefficient for the front velocity, $U_f\propto\sqrt{\kappa_0\Gamma_0}$, so there is in fact a single adjustable parameter. Furthermore, the thermal conduction exponent $\beta$, while still constrained within a certain range, also remains undefined. According to our previous work, its value modifies the internal structure of the deflagration front~\cite{Modestov_PRB_2011}, so we can expect nonlinear effects on the front velocity as well. Thus, having solved the dimensionless eigenvalue problem, we do not obtain actual velocity values but find the dependence of the front velocity as a function of the  magnetic field. The fitting parameters $\beta$ and $\kappa_0\Gamma_0$ can then be found by comparing the computed values to  experimental measurements.

\begin{figure*}
\centering
\includegraphics[width=6.6in]{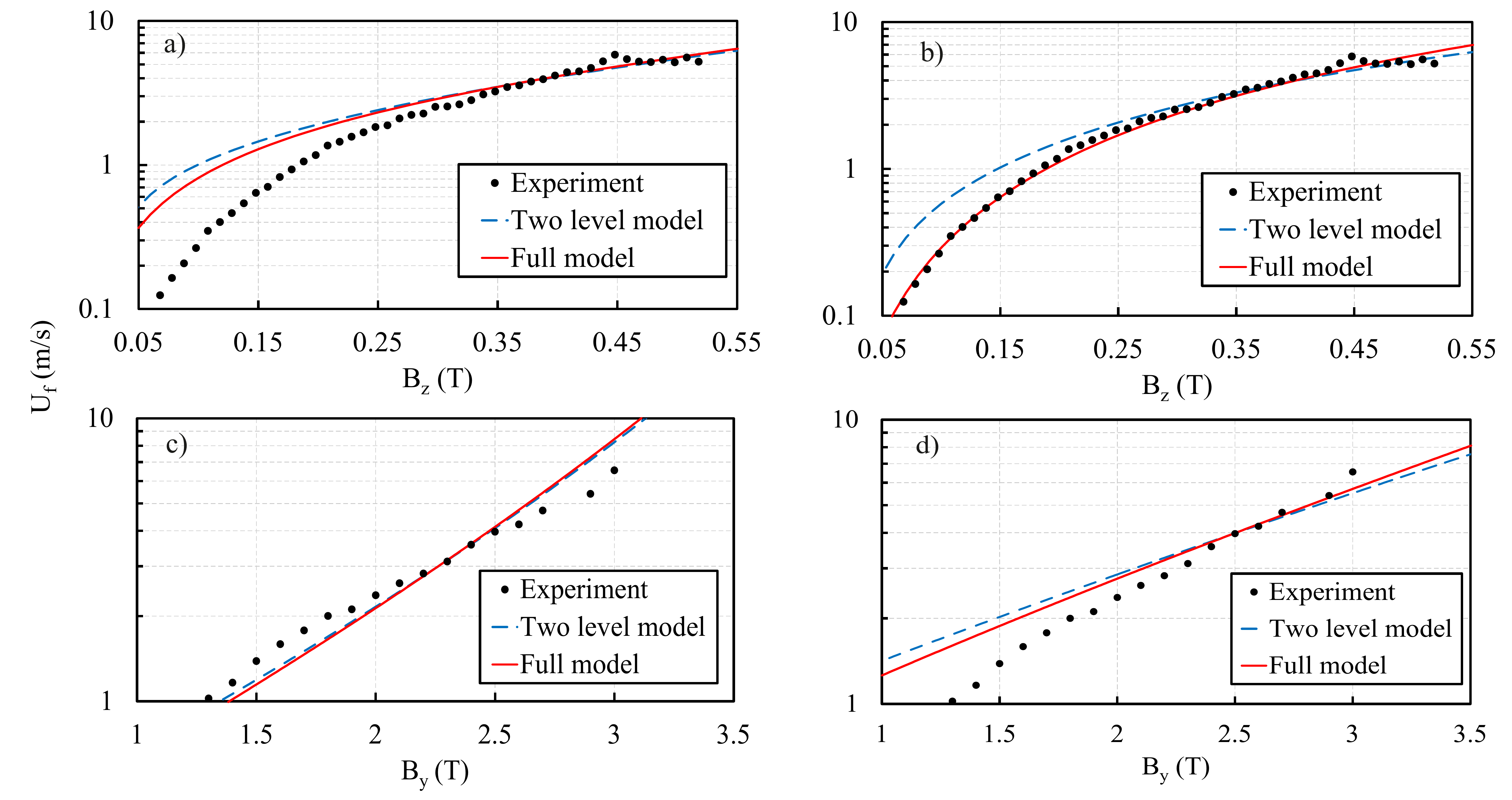}
\caption{\label{fig_UvsB}(color online) Magnetic deflagration velocity vs longitudinal (top panels) and transverse (bottom panels) magnetic fields for two thermal diffusivity exponents, $\beta=13/3$ (left panels) and $\beta=-3$ (right panels). Other parameters are: initial temperature $T_0 = \SI{0.4}{K}$; $B_y = \SI{2.5}{T}$  (for the upper panels) or $B_z = \SI{0.4}{T}$ (for the lower panels). The fitting coefficient  for the two-level model is $\Gamma_0\kappa_0 = \SI{2.6e5}{s K^{-13/3}}$ and for the full model is $\Gamma_0\kappa_0 = \SI{2.92e5}{s K^{-13/3}}$. The experimental data is taken from Ref.~\cite{Subedi_PRL_2013}.}
\end{figure*}

Generally speaking, all characteristic features of magnetic deflagration are governed by the external magnetic field (the effect of the initial temperature is vanishing and can be neglected). Consequently, the front velocity is regulated by both field components $B_z$ and $B_y$. In Fig.~\ref{fig_UvsB}, we plot the deflagration front velocity vs longitudinal (upper panels) and transverse (lower panels) magnetic fields for two values of the thermal conduction factor, $\beta=13/3$ and $\beta=-3$.
The solid line in Fig.~\ref{fig_UvsB} represents our full model, while the dashed line stands for simplified two-level model; experimental measurements for Ref.~\cite{Subedi_PRL_2013} are also depicted. Firstly, we notice that both theoretical models exhibit similar trends, in that the front velocity increases with the strength of the magnetic field.  The very similar quantitative behavior may appear to contradict the significant difference demonstrated in Fig.~\ref{rel_par}, but such a discrepancy can be explained by the relatively weak dependence of the velocity on the final temperature and energy release.  As a function of the longitudinal field, for $\beta=13/3$, Fig.~\ref{fig_UvsB}(a), both models predict a higher velocity, especially when the field is relatively weak; for stronger fields, the two models almost coincide with each other and with the experiment. For $\beta=-3$, Fig.~\ref{fig_UvsB}(b), the full model demonstrates perfect agreement with the experimental data over the entire range of the longitudinal magnetic field. A minor peak at $B_z \approx \SI{0.45}{T}$ is due to a quantum tunneling resonance that is not accounted for in the present theory. Also, it is of interest to note that decreasing $\beta$ leads to a steepening of the theoretical curves, showing a stronger dependence against the longitudinal magnetic field. The velocity dependence vs the transverse field for different $\beta$, Figs.~\ref{fig_UvsB}(c) and (d), shows the opposite trend. Moreover, there is here a better agreement with the experiment for $\beta=13/3$ than for $\beta=-3$.  It should be noted that the velocity plotted in Fig.~\ref{fig_UvsB} is on a logarithmic scale, which can visually obscure the difference between the two-level and the full models; for instance, in the case of Fig.~\ref{fig_UvsB}(b), the two-level model overestimates the velocity by a factor of up to 2.4.

We believe that this new model of magnetic deflagration describes the dependence of the front velocity $U_f$ on $B_z$ accurately enough, while its dependence on $B_y$ deserves additional study.  It is important to clarify in which way each of the magnetic field components affects the front velocity. In the governing equations there are two magnetic-field-dependent parameters which influence the front velocity: the Zeeman energy and the activation energy. Both of them depend on both field components. However, the Zeeman energy mostly depends on the longitudinal component of magnetic field; within the two-level model it is a linear function, $Q=2g\mu_B B_z S$. Consequently,  the dependence of the Zeeman energy (heat release) against the longitudinal field determines the corresponding relation for the front velocity as well. Hence, the more accurate calculation of the Zeeman energy developed in this paper results in a better agreement with the experimental data for the front velocity.

On the other hand, the activation energy is a complicated function of both components of the magnetic field. Moreover, in the presence of the transverse component, the simple double-well model~\cite{Gatteschi_ACIE_2003} with an activation energy may need to be reconsidered. As we see in Figs.~\ref{fig:states_compare} and \ref{fig:states3D}, the spin states do not follow a simple progression from one extreme to the other in the presence of a strong transverse field. This leads to an ambiguity in determining the activation energy as the highest-energy state the system must pass through during the spin flip becomes uncertain.  For instance, for $B_y=0$, all $2S+1$ states are aligned on the relaxation path (neglecting resonant tunneling), Fig.~\ref{fig:states3D}, from $-S$ to $S$, while for $B_y=3\ \textrm{T}$ only six levels appear to be on the relaxation path. Hence, by increasing the perpendicular field, the number of states involved in the relaxation process is reduced.  In addition, the transverse field also affects the spin-phonon coupling between adjacent states, which may also influence the front velocity.

\section{Conclusion}

In this paper, we have extended the previous theoretical model for the calculation of the heat release (Zeeman energy) during magnetic deflagration.  We have shown that in addition to metastable and ground states ($-S$ and $S$), other states also contribute to the energy release and must be taken into account. By building a theoretical model including all spin levels, we are able to take into account the partial spin flipping occurring in the nanomagnet crystal.

Using this new model, we have investigated influence of the transverse magnetic field on the deflagration front velocity.  We demonstrated that, due to thermal equilibrium populations, the higher spin levels may have more than 30\% total occupancy, leading to a significant difference in combustion temperature and front velocity.  Our model thus predicts a stronger dependence of the front velocity against the longitudinal magnetic field, showing a remarkable agreement with experimental measurements for a thermal diffusivity exponent of $\beta=-3$. 

However, the agreement of numerical simulations with experiment is less good in the presence of a strong transverse magnetic field.  We have shown that not all eigenstates of the Hamiltonian for a molecular magnet with both longitudinal and transverse magnetic fields participate in the spin relaxation process.  This makes the modeling of the magnetic deflagration using an Arrhenius law with a well-defined activation energy more difficult. We have nevertheless provided a classical model from which an activation energy can be calculated, that is in good agreement with the quantum mechanical level calculation. Future work should concentrate on refining the Arrhenius-law model in the presence of strong transverse magnetic fields.

\begin{acknowledgments}
  The authors thank Myriam Sarachik and Javier Tejada Palacios for useful discussions. Funding from the Swedish Research Council (VR) and the Faculty of Natural Sciences, Ume{\aa} University, is gratefully acknowledged.
\end{acknowledgments}

\bibliography{Multilevel}

\begin{thebibliography}{30}%
\makeatletter
\providecommand \@ifxundefined [1]{%
 \@ifx{#1\undefined}
}%
\providecommand \@ifnum [1]{%
 \ifnum #1\expandafter \@firstoftwo
 \else \expandafter \@secondoftwo
 \fi
}%
\providecommand \@ifx [1]{%
 \ifx #1\expandafter \@firstoftwo
 \else \expandafter \@secondoftwo
 \fi
}%
\providecommand \natexlab [1]{#1}%
\providecommand \enquote  [1]{``#1''}%
\providecommand \bibnamefont  [1]{#1}%
\providecommand \bibfnamefont [1]{#1}%
\providecommand \citenamefont [1]{#1}%
\providecommand \href@noop [0]{\@secondoftwo}%
\providecommand \href [0]{\begingroup \@sanitize@url \@href}%
\providecommand \@href[1]{\@@startlink{#1}\@@href}%
\providecommand \@@href[1]{\endgroup#1\@@endlink}%
\providecommand \@sanitize@url [0]{\catcode `\\12\catcode `\$12\catcode
  `\&12\catcode `\#12\catcode `\^12\catcode `\_12\catcode `\%12\relax}%
\providecommand \@@startlink[1]{}%
\providecommand \@@endlink[0]{}%
\providecommand \url  [0]{\begingroup\@sanitize@url \@url }%
\providecommand \@url [1]{\endgroup\@href {#1}{\urlprefix }}%
\providecommand \urlprefix  [0]{URL }%
\providecommand \Eprint [0]{\href }%
\providecommand \doibase [0]{http://dx.doi.org/}%
\providecommand \selectlanguage [0]{\@gobble}%
\providecommand \bibinfo  [0]{\@secondoftwo}%
\providecommand \bibfield  [0]{\@secondoftwo}%
\providecommand \translation [1]{[#1]}%
\providecommand \BibitemOpen [0]{}%
\providecommand \bibitemStop [0]{}%
\providecommand \bibitemNoStop [0]{.\EOS\space}%
\providecommand \EOS [0]{\spacefactor3000\relax}%
\providecommand \BibitemShut  [1]{\csname bibitem#1\endcsname}%
\let\auto@bib@innerbib\@empty
\bibitem [{\citenamefont {Sessoli}\ \emph
  {et~al.}(1993{\natexlab{a}})\citenamefont {Sessoli}, \citenamefont
  {Gatteschi}, \citenamefont {Caneschi},\ and\ \citenamefont
  {Novak}}]{Sessoli_Nature_1993}%
  \BibitemOpen
  \bibfield  {author} {\bibinfo {author} {\bibfnamefont {R.}~\bibnamefont
  {Sessoli}}, \bibinfo {author} {\bibfnamefont {D.}~\bibnamefont {Gatteschi}},
  \bibinfo {author} {\bibfnamefont {A.}~\bibnamefont {Caneschi}}, \ and\
  \bibinfo {author} {\bibfnamefont {M.~A.}\ \bibnamefont {Novak}},\ }\href@noop
  {} {\bibfield  {journal} {\bibinfo  {journal} {Nature}\ }\textbf {\bibinfo
  {volume} {365}},\ \bibinfo {pages} {141} (\bibinfo {year}
  {1993}{\natexlab{a}})}\BibitemShut {NoStop}%
\bibitem [{\citenamefont {Villain}\ \emph {et~al.}(1994)\citenamefont
  {Villain}, \citenamefont {Hartman-Boutron}, \citenamefont {Sessoli},\ and\
  \citenamefont {Rettori}}]{Villain_EPL_1994}%
  \BibitemOpen
  \bibfield  {author} {\bibinfo {author} {\bibfnamefont {J.}~\bibnamefont
  {Villain}}, \bibinfo {author} {\bibfnamefont {F.}~\bibnamefont
  {Hartman-Boutron}}, \bibinfo {author} {\bibfnamefont {R.}~\bibnamefont
  {Sessoli}}, \ and\ \bibinfo {author} {\bibfnamefont {A.}~\bibnamefont
  {Rettori}},\ }\href {http://stacks.iop.org/0295-5075/27/i=2/a=014} {\bibfield
   {journal} {\bibinfo  {journal} {Europhys. Lett.}\ }\textbf {\bibinfo
  {volume} {27}},\ \bibinfo {pages} {159} (\bibinfo {year} {1994})}\BibitemShut
  {NoStop}%
\bibitem [{\citenamefont {Friedman}\ \emph {et~al.}(1996)\citenamefont
  {Friedman}, \citenamefont {Sarachik}, \citenamefont {Tejada},\ and\
  \citenamefont {Ziolo}}]{Friedman_PRL_1996}%
  \BibitemOpen
  \bibfield  {author} {\bibinfo {author} {\bibfnamefont {J.~R.}\ \bibnamefont
  {Friedman}}, \bibinfo {author} {\bibfnamefont {M.~P.}\ \bibnamefont
  {Sarachik}}, \bibinfo {author} {\bibfnamefont {J.}~\bibnamefont {Tejada}}, \
  and\ \bibinfo {author} {\bibfnamefont {R.}~\bibnamefont {Ziolo}},\ }\href
  {\doibase 10.1103/PhysRevLett.76.3830} {\bibfield  {journal} {\bibinfo
  {journal} {Phys. Rev. Lett.}\ }\textbf {\bibinfo {volume} {76}},\ \bibinfo
  {pages} {3830} (\bibinfo {year} {1996})}\BibitemShut {NoStop}%
\bibitem [{\citenamefont {Thomas}\ \emph {et~al.}(1996)\citenamefont {Thomas},
  \citenamefont {Lionti}, \citenamefont {Ballou}, \citenamefont {Gatteschi},
  \citenamefont {Sessoli},\ and\ \citenamefont {Barbara}}]{Thomas_Nature_1996}%
  \BibitemOpen
  \bibfield  {author} {\bibinfo {author} {\bibfnamefont {L.}~\bibnamefont
  {Thomas}}, \bibinfo {author} {\bibfnamefont {F.}~\bibnamefont {Lionti}},
  \bibinfo {author} {\bibfnamefont {R.}~\bibnamefont {Ballou}}, \bibinfo
  {author} {\bibfnamefont {D.}~\bibnamefont {Gatteschi}}, \bibinfo {author}
  {\bibfnamefont {R.}~\bibnamefont {Sessoli}}, \ and\ \bibinfo {author}
  {\bibfnamefont {B.}~\bibnamefont {Barbara}},\ }\href@noop {} {\bibfield
  {journal} {\bibinfo  {journal} {Nature}\ }\textbf {\bibinfo {volume} {383}},\
  \bibinfo {pages} {145} (\bibinfo {year} {1996})}\BibitemShut {NoStop}%
\bibitem [{\citenamefont {Chudnovsky}\ and\ \citenamefont
  {Tejada}(1998)}]{Chudnovsky_Tejada_1998}%
  \BibitemOpen
  \bibfield  {author} {\bibinfo {author} {\bibfnamefont {E.~M.}\ \bibnamefont
  {Chudnovsky}}\ and\ \bibinfo {author} {\bibfnamefont {J.}~\bibnamefont
  {Tejada}},\ }\href@noop {} {\emph {\bibinfo {title} {Macroscopic Quantum
  Tunneling of the Magnetic Moment}}}\ (\bibinfo  {publisher} {Cambridge
  University Press},\ \bibinfo {address} {Cambridge,UK},\ \bibinfo {year}
  {1998})\BibitemShut {NoStop}%
\bibitem [{\citenamefont {Wernsdorfer}(2001)}]{Wernsdorfer_ACP_2001}%
  \BibitemOpen
  \bibfield  {author} {\bibinfo {author} {\bibfnamefont {W.}~\bibnamefont
  {Wernsdorfer}},\ }\href@noop {} {\bibfield  {journal} {\bibinfo  {journal}
  {Adv. Chem. Phys.}\ }\textbf {\bibinfo {volume} {118}},\ \bibinfo {pages}
  {99} (\bibinfo {year} {2001})}\BibitemShut {NoStop}%
\bibitem [{\citenamefont {Gatteschi}\ and\ \citenamefont
  {Sessoli}(2003)}]{Gatteschi_ACIE_2003}%
  \BibitemOpen
  \bibfield  {author} {\bibinfo {author} {\bibfnamefont {D.}~\bibnamefont
  {Gatteschi}}\ and\ \bibinfo {author} {\bibfnamefont {R.}~\bibnamefont
  {Sessoli}},\ }\href {\doibase 10.1002/anie.200390099} {\bibfield  {journal}
  {\bibinfo  {journal} {Angew. Chem. Int. Ed.}\ }\textbf {\bibinfo {volume}
  {42}},\ \bibinfo {pages} {268} (\bibinfo {year} {2003})}\BibitemShut
  {NoStop}%
\bibitem [{\citenamefont {del Barco}\ \emph {et~al.}(2005)\citenamefont {del
  Barco}, \citenamefont {Kent}, \citenamefont {Hill}, \citenamefont {North},
  \citenamefont {Dalal}, \citenamefont {Rumberger}, \citenamefont
  {Hendrickson}, \citenamefont {Chakov},\ and\ \citenamefont
  {Christou}}]{Delbarco_JLTP_2005}%
  \BibitemOpen
  \bibfield  {author} {\bibinfo {author} {\bibfnamefont {E.}~\bibnamefont {del
  Barco}}, \bibinfo {author} {\bibfnamefont {A.~D.}\ \bibnamefont {Kent}},
  \bibinfo {author} {\bibfnamefont {S.}~\bibnamefont {Hill}}, \bibinfo {author}
  {\bibfnamefont {J.~M.}\ \bibnamefont {North}}, \bibinfo {author}
  {\bibfnamefont {N.~S.}\ \bibnamefont {Dalal}}, \bibinfo {author}
  {\bibfnamefont {E.~M.}\ \bibnamefont {Rumberger}}, \bibinfo {author}
  {\bibfnamefont {D.~N.}\ \bibnamefont {Hendrickson}}, \bibinfo {author}
  {\bibfnamefont {N.}~\bibnamefont {Chakov}}, \ and\ \bibinfo {author}
  {\bibfnamefont {G.}~\bibnamefont {Christou}},\ }\href {\doibase
  10.1007/s10909-005-6016-3} {\bibfield  {journal} {\bibinfo  {journal} {J. Low
  Temp. Phys.}\ }\textbf {\bibinfo {volume} {140}},\ \bibinfo {pages} {119}
  (\bibinfo {year} {2005})}\BibitemShut {NoStop}%
\bibitem [{\citenamefont {Lis}(1980)}]{Lis_ACSB_1980}%
  \BibitemOpen
  \bibfield  {author} {\bibinfo {author} {\bibfnamefont {T.}~\bibnamefont
  {Lis}},\ }\href {\doibase 10.1107/S0567740880007893} {\bibfield  {journal}
  {\bibinfo  {journal} {Acta Crystallogr. Sect. B}\ }\textbf {\bibinfo {volume}
  {36}},\ \bibinfo {pages} {2042} (\bibinfo {year} {1980})}\BibitemShut
  {NoStop}%
\bibitem [{\citenamefont {Caneschi}\ \emph {et~al.}(1991)\citenamefont
  {Caneschi}, \citenamefont {Gatteschi}, \citenamefont {Sessoli}, \citenamefont
  {Barra}, \citenamefont {Brunel},\ and\ \citenamefont
  {Guillot}}]{Caneschi_JACS_1991}%
  \BibitemOpen
  \bibfield  {author} {\bibinfo {author} {\bibfnamefont {A.}~\bibnamefont
  {Caneschi}}, \bibinfo {author} {\bibfnamefont {D.}~\bibnamefont {Gatteschi}},
  \bibinfo {author} {\bibfnamefont {R.}~\bibnamefont {Sessoli}}, \bibinfo
  {author} {\bibfnamefont {A.~L.}\ \bibnamefont {Barra}}, \bibinfo {author}
  {\bibfnamefont {L.~C.}\ \bibnamefont {Brunel}}, \ and\ \bibinfo {author}
  {\bibfnamefont {M.}~\bibnamefont {Guillot}},\ }\href {\doibase
  10.1021/ja00015a057} {\bibfield  {journal} {\bibinfo  {journal} {J. Am. Chem.
  Soc.}\ }\textbf {\bibinfo {volume} {113}},\ \bibinfo {pages} {5873} (\bibinfo
  {year} {1991})}\BibitemShut {NoStop}%
\bibitem [{\citenamefont {Papaefthymiou}(1992)}]{Papaefthymiou_PRB_1992}%
  \BibitemOpen
  \bibfield  {author} {\bibinfo {author} {\bibfnamefont {G.~C.}\ \bibnamefont
  {Papaefthymiou}},\ }\href {\doibase 10.1103/PhysRevB.46.10366} {\bibfield
  {journal} {\bibinfo  {journal} {Phys. Rev. B}\ }\textbf {\bibinfo {volume}
  {46}},\ \bibinfo {pages} {10366} (\bibinfo {year} {1992})}\BibitemShut
  {NoStop}%
\bibitem [{\citenamefont {Sessoli}\ \emph
  {et~al.}(1993{\natexlab{b}})\citenamefont {Sessoli}, \citenamefont {Tsai},
  \citenamefont {Schake}, \citenamefont {Wang}, \citenamefont {Vincent},
  \citenamefont {Folting}, \citenamefont {Gatteschi}, \citenamefont
  {Christou},\ and\ \citenamefont {Hendrickson}}]{Sessoli_JACS_1993}%
  \BibitemOpen
  \bibfield  {author} {\bibinfo {author} {\bibfnamefont {R.}~\bibnamefont
  {Sessoli}}, \bibinfo {author} {\bibfnamefont {H.~L.}\ \bibnamefont {Tsai}},
  \bibinfo {author} {\bibfnamefont {A.~R.}\ \bibnamefont {Schake}}, \bibinfo
  {author} {\bibfnamefont {S.}~\bibnamefont {Wang}}, \bibinfo {author}
  {\bibfnamefont {J.~B.}\ \bibnamefont {Vincent}}, \bibinfo {author}
  {\bibfnamefont {K.}~\bibnamefont {Folting}}, \bibinfo {author} {\bibfnamefont
  {D.}~\bibnamefont {Gatteschi}}, \bibinfo {author} {\bibfnamefont
  {G.}~\bibnamefont {Christou}}, \ and\ \bibinfo {author} {\bibfnamefont
  {D.~N.}\ \bibnamefont {Hendrickson}},\ }\href {\doibase 10.1021/ja00058a027}
  {\bibfield  {journal} {\bibinfo  {journal} {J. Am. Chem. Soc.}\ }\textbf
  {\bibinfo {volume} {115}},\ \bibinfo {pages} {1804} (\bibinfo {year}
  {1993}{\natexlab{b}})}\BibitemShut {NoStop}%
\bibitem [{\citenamefont {Bhanja}\ \emph {et~al.}(2016)\citenamefont {Bhanja},
  \citenamefont {Karunaratne}, \citenamefont {Panchumarthy},\ and\
  \citenamefont {Srinath~Rajaram}}]{Bhanja_NatNano_2015}%
  \BibitemOpen
  \bibfield  {author} {\bibinfo {author} {\bibfnamefont {S.}~\bibnamefont
  {Bhanja}}, \bibinfo {author} {\bibfnamefont {D.~K.}\ \bibnamefont
  {Karunaratne}}, \bibinfo {author} {\bibfnamefont {R.}~\bibnamefont
  {Panchumarthy}}, \ and\ \bibinfo {author} {\bibfnamefont {S.~S.}\
  \bibnamefont {Srinath~Rajaram}},\ }\href {\doibase 10.1038/nnano.2015.245}
  {\bibfield  {journal} {\bibinfo  {journal} {Nature Nanotech.}\ }\textbf
  {\bibinfo {volume} {11}},\ \bibinfo {pages} {177–183} (\bibinfo {year}
  {2016})}\BibitemShut {NoStop}%
\bibitem [{\citenamefont {Suzuki}\ \emph {et~al.}(2005)\citenamefont {Suzuki},
  \citenamefont {Sarachik}, \citenamefont {Chudnovsky}, \citenamefont {McHugh},
  \citenamefont {Gonzalez-Rubio}, \citenamefont {Avraham}, \citenamefont
  {Myasoedov}, \citenamefont {Zeldov}, \citenamefont {Shtrikman}, \citenamefont
  {Chakov},\ and\ \citenamefont {Christou}}]{Suzuki_PRL_2005}%
  \BibitemOpen
  \bibfield  {author} {\bibinfo {author} {\bibfnamefont {Y.}~\bibnamefont
  {Suzuki}}, \bibinfo {author} {\bibfnamefont {M.~P.}\ \bibnamefont
  {Sarachik}}, \bibinfo {author} {\bibfnamefont {E.~M.}\ \bibnamefont
  {Chudnovsky}}, \bibinfo {author} {\bibfnamefont {S.}~\bibnamefont {McHugh}},
  \bibinfo {author} {\bibfnamefont {R.}~\bibnamefont {Gonzalez-Rubio}},
  \bibinfo {author} {\bibfnamefont {N.}~\bibnamefont {Avraham}}, \bibinfo
  {author} {\bibfnamefont {Y.}~\bibnamefont {Myasoedov}}, \bibinfo {author}
  {\bibfnamefont {E.}~\bibnamefont {Zeldov}}, \bibinfo {author} {\bibfnamefont
  {H.}~\bibnamefont {Shtrikman}}, \bibinfo {author} {\bibfnamefont {N.~E.}\
  \bibnamefont {Chakov}}, \ and\ \bibinfo {author} {\bibfnamefont
  {G.}~\bibnamefont {Christou}},\ }\href {\doibase
  10.1103/PhysRevLett.95.147201} {\bibfield  {journal} {\bibinfo  {journal}
  {Phys. Rev. Lett.}\ }\textbf {\bibinfo {volume} {95}},\ \bibinfo {pages}
  {147201} (\bibinfo {year} {2005})}\BibitemShut {NoStop}%
\bibitem [{\citenamefont {Hernández-Mínguez}\ \emph
  {et~al.}(2005)\citenamefont {Hernández-Mínguez}, \citenamefont {Hernandez},
  \citenamefont {Maci\`a}, \citenamefont {Garc{\'\i}a-Santiago}, \citenamefont
  {Tejada},\ and\ \citenamefont {Santos}}]{Hernandez-Minguez_PRL_2005}%
  \BibitemOpen
  \bibfield  {author} {\bibinfo {author} {\bibfnamefont {A.}~\bibnamefont
  {Hernández-Mínguez}}, \bibinfo {author} {\bibfnamefont {J.~M.}\
  \bibnamefont {Hernandez}}, \bibinfo {author} {\bibfnamefont {F.}~\bibnamefont
  {Maci\`a}}, \bibinfo {author} {\bibfnamefont {A.}~\bibnamefont
  {Garc{\'\i}a-Santiago}}, \bibinfo {author} {\bibfnamefont {J.}~\bibnamefont
  {Tejada}}, \ and\ \bibinfo {author} {\bibfnamefont {P.~V.}\ \bibnamefont
  {Santos}},\ }\href {\doibase 10.1103/PhysRevLett.95.217205} {\bibfield
  {journal} {\bibinfo  {journal} {Phys. Rev. Lett.}\ }\textbf {\bibinfo
  {volume} {95}},\ \bibinfo {pages} {217205} (\bibinfo {year}
  {2005})}\BibitemShut {NoStop}%
\bibitem [{\citenamefont {Garanin}\ and\ \citenamefont
  {Chudnovsky}(2007)}]{Garanin_PRB_2007}%
  \BibitemOpen
  \bibfield  {author} {\bibinfo {author} {\bibfnamefont {D.~A.}\ \bibnamefont
  {Garanin}}\ and\ \bibinfo {author} {\bibfnamefont {E.~M.}\ \bibnamefont
  {Chudnovsky}},\ }\href {\doibase 10.1103/PhysRevB.76.054410} {\bibfield
  {journal} {\bibinfo  {journal} {Phys. Rev. B}\ }\textbf {\bibinfo {volume}
  {76}},\ \bibinfo {pages} {054410} (\bibinfo {year} {2007})}\BibitemShut
  {NoStop}%
\bibitem [{\citenamefont {Villuendas}\ \emph {et~al.}(2008)\citenamefont
  {Villuendas}, \citenamefont {Gheorghe}, \citenamefont
  {Hern\'andez-M{\'\i}nguez}, \citenamefont {Maci\`a}, \citenamefont
  {Hernandez}, \citenamefont {Tejada},\ and\ \citenamefont
  {Wijngaarden}}]{Villuendas_EPL_2008}%
  \BibitemOpen
  \bibfield  {author} {\bibinfo {author} {\bibfnamefont {D.}~\bibnamefont
  {Villuendas}}, \bibinfo {author} {\bibfnamefont {D.}~\bibnamefont
  {Gheorghe}}, \bibinfo {author} {\bibfnamefont {A.}~\bibnamefont
  {Hern\'andez-M{\'\i}nguez}}, \bibinfo {author} {\bibfnamefont
  {F.}~\bibnamefont {Maci\`a}}, \bibinfo {author} {\bibfnamefont {J.~M.}\
  \bibnamefont {Hernandez}}, \bibinfo {author} {\bibfnamefont {J.}~\bibnamefont
  {Tejada}}, \ and\ \bibinfo {author} {\bibfnamefont {R.~J.}\ \bibnamefont
  {Wijngaarden}},\ }\href {http://stacks.iop.org/0295-5075/84/i=6/a=67010}
  {\bibfield  {journal} {\bibinfo  {journal} {EPL}\ }\textbf {\bibinfo {volume}
  {84}},\ \bibinfo {pages} {67010} (\bibinfo {year} {2008})}\BibitemShut
  {NoStop}%
\bibitem [{\citenamefont {Decelle}\ \emph {et~al.}(2009)\citenamefont
  {Decelle}, \citenamefont {Vanacken}, \citenamefont {Moshchalkov},
  \citenamefont {Tejada}, \citenamefont {Hern\'andez},\ and\ \citenamefont
  {Maci\`a}}]{Decelle_PRL_2009}%
  \BibitemOpen
  \bibfield  {author} {\bibinfo {author} {\bibfnamefont {W.}~\bibnamefont
  {Decelle}}, \bibinfo {author} {\bibfnamefont {J.}~\bibnamefont {Vanacken}},
  \bibinfo {author} {\bibfnamefont {V.~V.}\ \bibnamefont {Moshchalkov}},
  \bibinfo {author} {\bibfnamefont {J.}~\bibnamefont {Tejada}}, \bibinfo
  {author} {\bibfnamefont {J.~M.}\ \bibnamefont {Hern\'andez}}, \ and\ \bibinfo
  {author} {\bibfnamefont {F.}~\bibnamefont {Maci\`a}},\ }\href {\doibase
  10.1103/PhysRevLett.102.027203} {\bibfield  {journal} {\bibinfo  {journal}
  {Phys. Rev. Lett.}\ }\textbf {\bibinfo {volume} {102}},\ \bibinfo {pages}
  {027203} (\bibinfo {year} {2009})}\BibitemShut {NoStop}%
\bibitem [{\citenamefont {Modestov}\ \emph {et~al.}(2011)\citenamefont
  {Modestov}, \citenamefont {Bychkov},\ and\ \citenamefont
  {Marklund}}]{Modestov_PRB_2011}%
  \BibitemOpen
  \bibfield  {author} {\bibinfo {author} {\bibfnamefont {M.}~\bibnamefont
  {Modestov}}, \bibinfo {author} {\bibfnamefont {V.}~\bibnamefont {Bychkov}}, \
  and\ \bibinfo {author} {\bibfnamefont {M.}~\bibnamefont {Marklund}},\ }\href
  {\doibase 10.1103/PhysRevB.83.214417} {\bibfield  {journal} {\bibinfo
  {journal} {Phys. Rev. B}\ }\textbf {\bibinfo {volume} {83}},\ \bibinfo
  {pages} {214417} (\bibinfo {year} {2011})}\BibitemShut {NoStop}%
\bibitem [{\citenamefont {Law}(2006)}]{Law-book}%
  \BibitemOpen
  \bibfield  {author} {\bibinfo {author} {\bibfnamefont {C.~K.}\ \bibnamefont
  {Law}},\ }\href@noop {} {\emph {\bibinfo {title} {Combustion Physics}}}\
  (\bibinfo  {publisher} {Cambridge University Press},\ \bibinfo {address}
  {Cambridge},\ \bibinfo {year} {2006})\BibitemShut {NoStop}%
\bibitem [{\citenamefont {Bychkov}\ and\ \citenamefont
  {Liberman}(2000)}]{Bychkov_PR_2000}%
  \BibitemOpen
  \bibfield  {author} {\bibinfo {author} {\bibfnamefont {V.}~\bibnamefont
  {Bychkov}}\ and\ \bibinfo {author} {\bibfnamefont {M.}~\bibnamefont
  {Liberman}},\ }\href {\doibase 10.1016/S0370-1573(99)00081-2} {\bibfield
  {journal} {\bibinfo  {journal} {Phys. Rep.}\ }\textbf {\bibinfo {volume}
  {325}},\ \bibinfo {pages} {115 } (\bibinfo {year} {2000})}\BibitemShut
  {NoStop}%
\bibitem [{\citenamefont {Friedman}(1998)}]{Friedman_PRB_98}%
  \BibitemOpen
  \bibfield  {author} {\bibinfo {author} {\bibfnamefont {J.~R.}\ \bibnamefont
  {Friedman}},\ }\href@noop {} {\bibfield  {journal} {\bibinfo  {journal}
  {Phys. Rev. B}\ }\textbf {\bibinfo {volume} {57}},\ \bibinfo {pages} {10291}
  (\bibinfo {year} {1998})}\BibitemShut {NoStop}%
\bibitem [{\citenamefont {Dion}\ \emph {et~al.}(2013)\citenamefont {Dion},
  \citenamefont {Jukimenko}, \citenamefont {Modestov}, \citenamefont
  {Marklund},\ and\ \citenamefont {Bychkov}}]{Dion_PRB_2013}%
  \BibitemOpen
  \bibfield  {author} {\bibinfo {author} {\bibfnamefont {C.~M.}\ \bibnamefont
  {Dion}}, \bibinfo {author} {\bibfnamefont {O.}~\bibnamefont {Jukimenko}},
  \bibinfo {author} {\bibfnamefont {M.}~\bibnamefont {Modestov}}, \bibinfo
  {author} {\bibfnamefont {M.}~\bibnamefont {Marklund}}, \ and\ \bibinfo
  {author} {\bibfnamefont {V.}~\bibnamefont {Bychkov}},\ }\href {\doibase
  10.1103/PhysRevB.87.014409} {\bibfield  {journal} {\bibinfo  {journal} {Phys.
  Rev. B}\ }\textbf {\bibinfo {volume} {87}},\ \bibinfo {pages} {014409}
  (\bibinfo {year} {2013})}\BibitemShut {NoStop}%
\bibitem [{\citenamefont {Subedi}\ \emph {et~al.}(2013)\citenamefont {Subedi},
  \citenamefont {V\'elez}, \citenamefont {Maci\`a}, \citenamefont {Li},
  \citenamefont {Sarachik}, \citenamefont {Tejada}, \citenamefont {Mukherjee},
  \citenamefont {Christou},\ and\ \citenamefont {Kent}}]{Subedi_PRL_2013}%
  \BibitemOpen
  \bibfield  {author} {\bibinfo {author} {\bibfnamefont {P.}~\bibnamefont
  {Subedi}}, \bibinfo {author} {\bibfnamefont {S.}~\bibnamefont {V\'elez}},
  \bibinfo {author} {\bibfnamefont {F.}~\bibnamefont {Maci\`a}}, \bibinfo
  {author} {\bibfnamefont {S.}~\bibnamefont {Li}}, \bibinfo {author}
  {\bibfnamefont {M.~P.}\ \bibnamefont {Sarachik}}, \bibinfo {author}
  {\bibfnamefont {J.}~\bibnamefont {Tejada}}, \bibinfo {author} {\bibfnamefont
  {S.}~\bibnamefont {Mukherjee}}, \bibinfo {author} {\bibfnamefont
  {G.}~\bibnamefont {Christou}}, \ and\ \bibinfo {author} {\bibfnamefont
  {A.~D.}\ \bibnamefont {Kent}},\ }\href {\doibase
  10.1103/PhysRevLett.110.207203} {\bibfield  {journal} {\bibinfo  {journal}
  {Phys. Rev. Lett.}\ }\textbf {\bibinfo {volume} {110}},\ \bibinfo {pages}
  {207203} (\bibinfo {year} {2013})}\BibitemShut {NoStop}%
\bibitem [{\citenamefont {V\'elez}\ \emph {et~al.}(2014)\citenamefont
  {V\'elez}, \citenamefont {Subedi}, \citenamefont {Maci\`a}, \citenamefont
  {Li}, \citenamefont {Sarachik}, \citenamefont {Tejada}, \citenamefont
  {Mukherjee}, \citenamefont {Christou},\ and\ \citenamefont
  {Kent}}]{Velez_PRB_2014}%
  \BibitemOpen
  \bibfield  {author} {\bibinfo {author} {\bibfnamefont {S.}~\bibnamefont
  {V\'elez}}, \bibinfo {author} {\bibfnamefont {P.}~\bibnamefont {Subedi}},
  \bibinfo {author} {\bibfnamefont {F.}~\bibnamefont {Maci\`a}}, \bibinfo
  {author} {\bibfnamefont {S.}~\bibnamefont {Li}}, \bibinfo {author}
  {\bibfnamefont {M.~P.}\ \bibnamefont {Sarachik}}, \bibinfo {author}
  {\bibfnamefont {J.}~\bibnamefont {Tejada}}, \bibinfo {author} {\bibfnamefont
  {S.}~\bibnamefont {Mukherjee}}, \bibinfo {author} {\bibfnamefont
  {G.}~\bibnamefont {Christou}}, \ and\ \bibinfo {author} {\bibfnamefont
  {A.~D.}\ \bibnamefont {Kent}},\ }\href {\doibase 10.1103/PhysRevB.89.144408}
  {\bibfield  {journal} {\bibinfo  {journal} {Phys. Rev. B}\ }\textbf {\bibinfo
  {volume} {89}},\ \bibinfo {pages} {144408} (\bibinfo {year}
  {2014})}\BibitemShut {NoStop}%
\bibitem [{\citenamefont {Garanin}(2013)}]{Garanin_PRB_2013}%
  \BibitemOpen
  \bibfield  {author} {\bibinfo {author} {\bibfnamefont {D.~A.}\ \bibnamefont
  {Garanin}},\ }\href {\doibase 10.1103/PhysRevB.88.064413} {\bibfield
  {journal} {\bibinfo  {journal} {Phys. Rev. B}\ }\textbf {\bibinfo {volume}
  {88}},\ \bibinfo {pages} {064413} (\bibinfo {year} {2013})}\BibitemShut
  {NoStop}%
\bibitem [{\citenamefont {Jukimenko}\ \emph {et~al.}(2014)\citenamefont
  {Jukimenko}, \citenamefont {Dion}, \citenamefont {Marklund},\ and\
  \citenamefont {Bychkov}}]{Jukimenko_PRL_2014}%
  \BibitemOpen
  \bibfield  {author} {\bibinfo {author} {\bibfnamefont {O.}~\bibnamefont
  {Jukimenko}}, \bibinfo {author} {\bibfnamefont {C.~M.}\ \bibnamefont {Dion}},
  \bibinfo {author} {\bibfnamefont {M.}~\bibnamefont {Marklund}}, \ and\
  \bibinfo {author} {\bibfnamefont {V.}~\bibnamefont {Bychkov}},\ }\href
  {\doibase 10.1103/PhysRevLett.113.217206} {\bibfield  {journal} {\bibinfo
  {journal} {Phys. Rev. Lett.}\ }\textbf {\bibinfo {volume} {113}},\ \bibinfo
  {pages} {217206} (\bibinfo {year} {2014})}\BibitemShut {NoStop}%
\bibitem [{\citenamefont {Garanin}\ and\ \citenamefont
  {Shoyeb}(2012)}]{Garanin_PRB_2012}%
  \BibitemOpen
  \bibfield  {author} {\bibinfo {author} {\bibfnamefont {D.~A.}\ \bibnamefont
  {Garanin}}\ and\ \bibinfo {author} {\bibfnamefont {S.}~\bibnamefont
  {Shoyeb}},\ }\href {\doibase 10.1103/PhysRevB.85.094403} {\bibfield
  {journal} {\bibinfo  {journal} {Phys. Rev. B}\ }\textbf {\bibinfo {volume}
  {85}},\ \bibinfo {pages} {094403} (\bibinfo {year} {2012})}\BibitemShut
  {NoStop}%
\bibitem [{\citenamefont {Koloskova}(1963)}]{Koloskova_FTT_1963}%
  \BibitemOpen
  \bibfield  {author} {\bibinfo {author} {\bibfnamefont {N.~G.}\ \bibnamefont
  {Koloskova}},\ }\href@noop {} {\bibfield  {journal} {\bibinfo  {journal}
  {Fiz. Tverd. Tela}\ }\textbf {\bibinfo {volume} {5}},\ \bibinfo {pages} {61}
  (\bibinfo {year} {1963})},\ \bibinfo {note} {[English transl.: \emph{Soviet
  Phys. -- Solid State} \textbf{5}, 40 (1963)]}\BibitemShut {NoStop}%
\bibitem [{\citenamefont {McHugh}\ \emph {et~al.}(2009)\citenamefont {McHugh},
  \citenamefont {Wen}, \citenamefont {Ma}, \citenamefont {Sarachik},
  \citenamefont {Myasoedov}, \citenamefont {Zeldov}, \citenamefont {Bagai},\
  and\ \citenamefont {Christou}}]{McHugh_PRB_2009}%
  \BibitemOpen
  \bibfield  {author} {\bibinfo {author} {\bibfnamefont {S.}~\bibnamefont
  {McHugh}}, \bibinfo {author} {\bibfnamefont {B.}~\bibnamefont {Wen}},
  \bibinfo {author} {\bibfnamefont {X.}~\bibnamefont {Ma}}, \bibinfo {author}
  {\bibfnamefont {M.}~\bibnamefont {Sarachik}}, \bibinfo {author}
  {\bibfnamefont {Y.}~\bibnamefont {Myasoedov}}, \bibinfo {author}
  {\bibfnamefont {E.}~\bibnamefont {Zeldov}}, \bibinfo {author} {\bibfnamefont
  {R.}~\bibnamefont {Bagai}}, \ and\ \bibinfo {author} {\bibfnamefont
  {G.}~\bibnamefont {Christou}},\ }\href
  {http://www.scopus.com/inward/record.url?eid=2-s2.0-66849119092&partnerID=40&md5=63942b158f8332d9ee6ffac566e845bf}
  {\bibfield  {journal} {\bibinfo  {journal} {Phys. Rev. B}\ }\textbf {\bibinfo
  {volume} {79}} (\bibinfo {year} {2009})}\BibitemShut {NoStop}%
\end{thebibliography}%

\end{document}